\documentclass[runningheads]{llncs}
\usepackage[T1]{fontenc}
\usepackage{graphicx}
\usepackage{subcaption}
\usepackage{amsmath}
\usepackage{amssymb}
\usepackage{booktabs}
\usepackage{booktabs} % For better looking tables
\usepackage{caption}  % For caption and label control
\usepackage{lipsum}   % For dummy text (optional)
\usepackage[colorlinks=true,allcolors=blue]{hyperref}
\begin{document}
\title{Towards Interpretable Radiology Report Generation via Concept Bottlenecks using a Multi-Agentic RAG}
\titlerunning{Towards Interpretable Radiology Report ...}

\author{Hasan Md Tusfiqur Alam\inst{1}\thanks{Corresponding author. Email: hasan.alam@dfki.de}\orcidID{0000-0003-1479-7690}
\and
Devansh Srivastav\inst{1}\orcidID{0000-0002-8858-7402}
\and 
Md Abdul Kadir\inst{1,2}\orcidID{0000-0002-8420-2536} \and
Daniel Sonntag\inst{1,2}\orcidID{0000-0002-8857-8709}}
\authorrunning{Alam et al.}
\institute{German Research Center for Artificial Intelligence (DFKI), Saarbrücken, Germany
\and
University of Oldenburg, Oldenburg, Germany
\email{\{hasan.alam,devansh.srivastav,abdul.kadir,daniel.sonntag\}@dfki.de}}
\maketitle              % typeset the header of the contribution
\begin{abstract}
Deep learning has advanced medical image classification, but interpretability challenges hinder its clinical adoption. This study enhances interpretability in Chest X-ray (CXR) classification by using concept bottleneck models (CBMs) and a multi-agent Retrieval-Augmented Generation (RAG) system for report generation. By modeling relationships between visual features and clinical concepts, we create interpretable concept vectors that guide a multi-agent RAG system to generate radiology reports, enhancing clinical relevance, explainability, and transparency. Evaluation of the generated reports using an LLM-as-a-judge confirmed the interpretability and clinical utility of our model’s outputs. On the COVID-QU dataset, our model achieved 81\% classification accuracy and demonstrated robust report generation performance, with five key metrics ranging between 84\% and 90\%. This interpretable multi-agent framework bridges the gap between high-performance AI and the explainability required for reliable AI-driven CXR analysis in clinical settings. Our code will be released at \href{https://github.com/tifat58/IRR-with-CBM-RAG.git}{https://github.com/tifat58/IRR-with-CBM-RAG}


\end{abstract}

\section{Introduction and Related Work}

Deep learning has significantly improved diagnostic accuracy in medical imaging, especially in chest X-ray (CXR) analysis for conditions such as pneumonia, lung cancer, and tuberculosis \cite{srivastav2021improved,hwang2019development}. However, these models often function as "\textit{black boxes}", limiting interpretability and clinician trust in critical fields like radiology \cite{de2023classification}. Traditional CXR classification methods lack transparency, providing little insight into model predictions. Although automated radiology report generation could streamline diagnostics and improve reporting consistency \cite{jing2017automatic}, it remains challenging to interpret and validate, which hinders its adoption in clinical settings that demand explainability \cite{holzinger2019causability}.

\begin{figure}[t]
    \centering
    \includegraphics[width=0.95\linewidth]{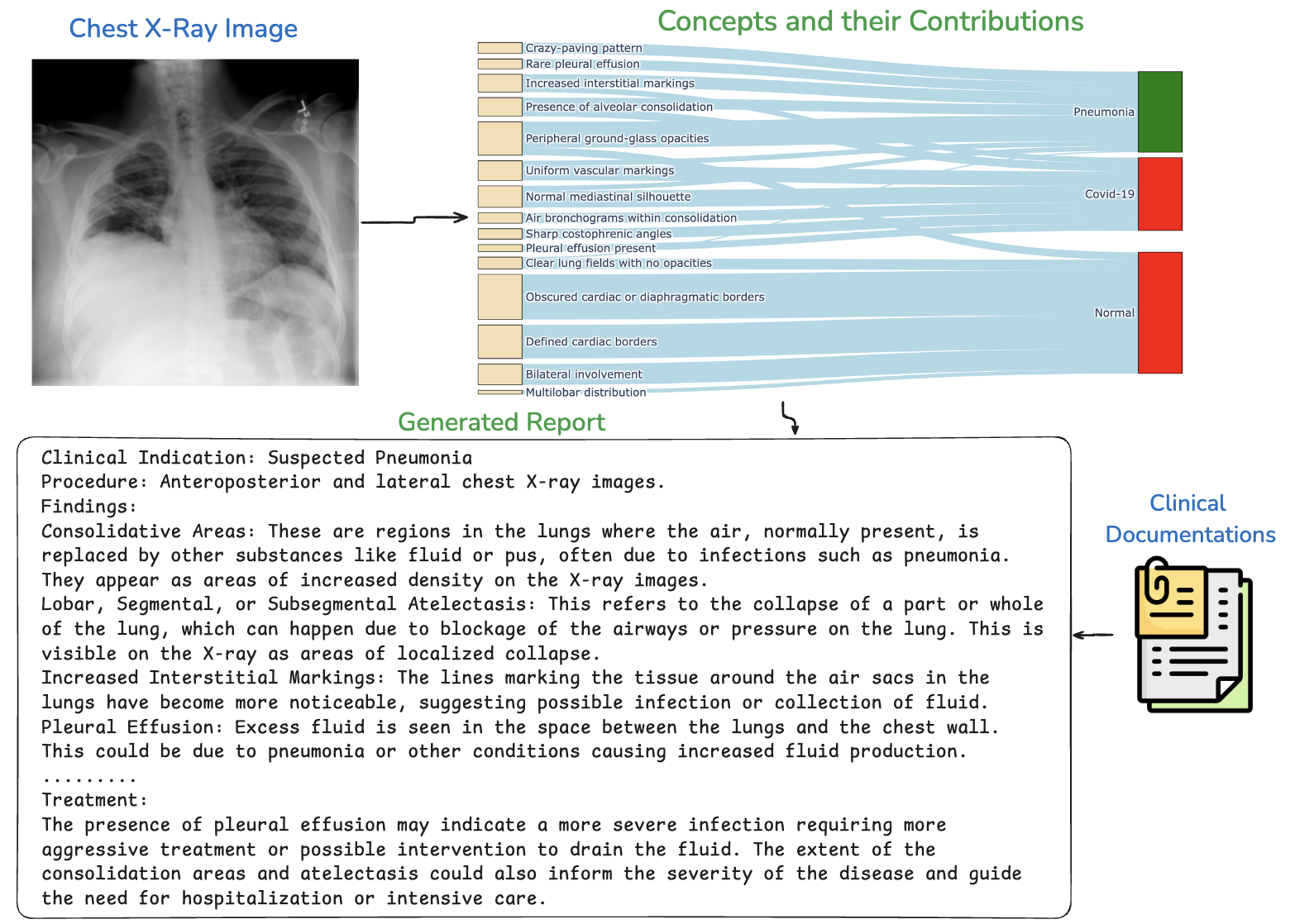}
    \caption{Input (\textit{blue}) to output (\textit{green}) pipeline: 
    Given a CXR as input, the Concept Bottleneck Model predicts clinical attributes (concepts) and their contributions in an intermediate step, followed by predicting the disease class. The multi-agent RAG system then generates a comprehensive report, incorporating clinical interpretations and insights drawn from relevant clinical documents}
    \label{fig:workflow}
\end{figure}

To address these challenges, we combine concept bottleneck models (CBMs) \cite{koh2020concept} with a multi-agent Retrieval-Augmented Generation (RAG) \cite{lewis2020retrieval,li2024more} system to enhance interpretability in CXR classification and report generation (Fig. \ref{fig:workflow}). While RAG systems have shown promise for radiology report generation by improving factual accuracy and reducing irrelevant content \cite{ranjit2023retrieval,bernardi2024report,sun2024fact,xia2024mmed,liang2024optimizing}, they face limitations: retrieval-only methods can introduce noise or redundancy, and generative models risk producing clinically inconsistent outputs. Our approach addresses these issues by modeling relationships between visual features and clinical concepts, creating interpretable concept vectors that support accurate classification and clear, clinically relevant reporting. 

CBMs enable interpretability by introducing concept layers where each neuron corresponds to a human-interpretable clinical concept \cite{koh2020concept,chauhan2023interactive}. Although traditional CBMs require extensive human annotation, which limits scalability, recent adaptations leverage vision-language models like CLIP \cite{radford2021learning} to align visual features with concepts \cite{oikarinen2023label,yuksekgonul2022post}. While these models typically incorporate complex components like residual connections and large concept sets, recent work by \cite{yan2023robust} suggests that simpler designs may be more effective for medical data, achieving robust representation without extensive modifications.
By bridging the gap between AI models and the interpretability necessary for effective medical decision-making, our approach aims to build an interpretable CXR interpretation system that empowers radiologists with greater insight into the AI’s decision-making process, supporting more informed and trustworthy diagnoses.

\section{Methodology}

\begin{figure}[t]
    \centering
    \includegraphics[width=0.92\linewidth]{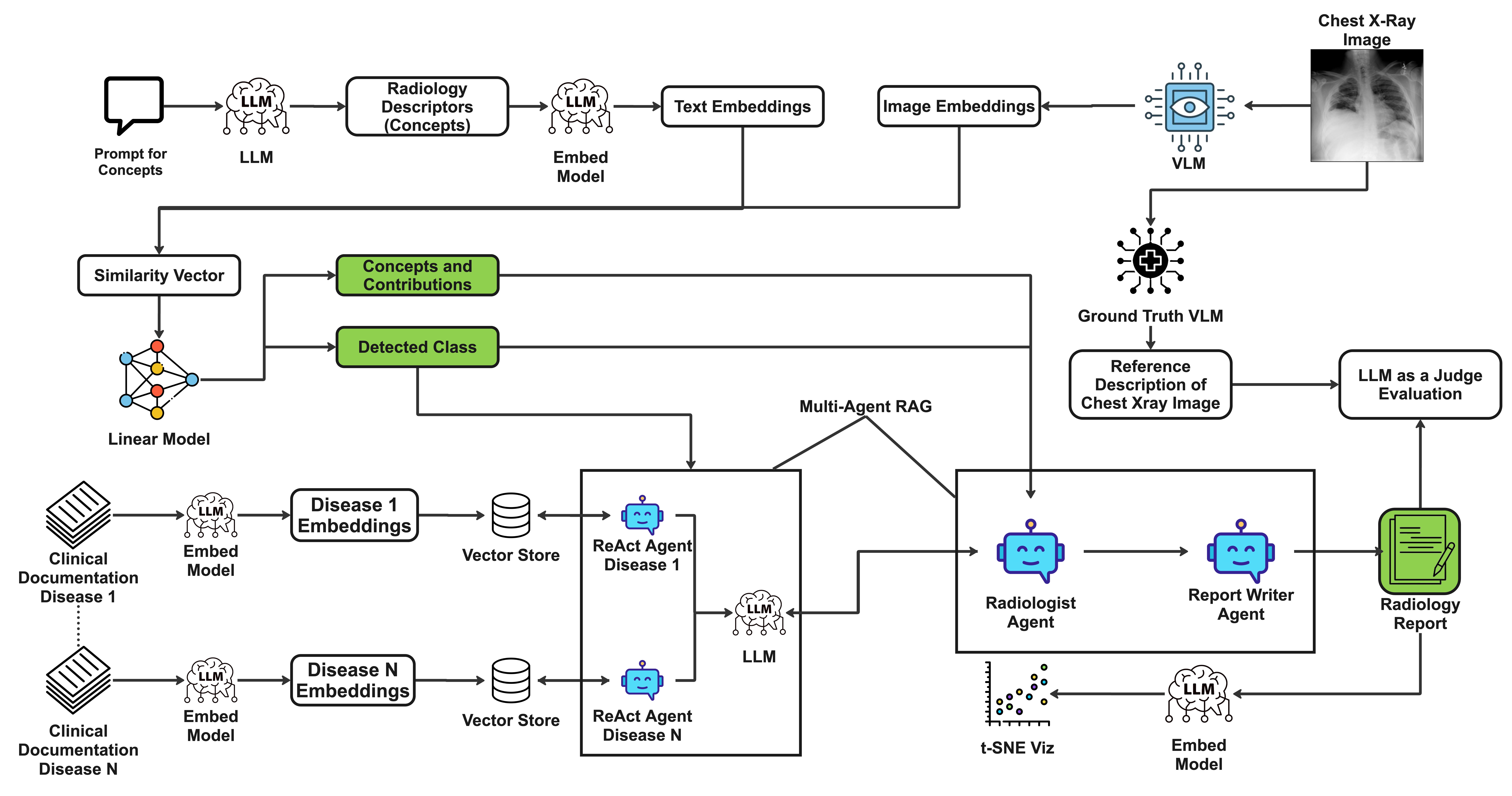}
    \caption{Proposed architecture for the interpretable report generation. (\textit{Top}) For a CXR image, disease class, and concept contribution scores are predicted using a CBM model with automatic concept discovery. (\textit{Bottom}) Based on these contributions, a multi-agent RAG system generates reports using relevant clinical documents. The chain-of-thought reasoning ensures that detected features contribute to accurate classification and report generation, with the final output evaluated for robustness and clinical relevance by LLM as judge \cite{zheng2023judging} evaluation.}
    \label{arch}
\end{figure}

Our approach leverages concept bottleneck models (CBMs) \cite{koh2020concept}, medical image embeddings and multi-agent RAG \cite{lewis2020retrieval} to ensure robustness and interpretability in medical image classification and report generation as shown in Fig. \ref{arch}, in two stages. First, disease classification with associated concept contributions (sec. \ref{subsec:cls}). Second, robust report generation using relevant clinical documents and descriptors from stage 1 (sec. \ref{subsec:rg}).

\subsection{Interpretable Classification using Concept Bottleneck}
\label{subsec:cls}
Following the \cite{oikarinen2023label}, we adopt CBMs with automatic concept discovery for CXR classification. We use GPT-4 \cite{achiam2023gpt} to query a set of $N = 20$ medical descriptors (concepts) for each disease category following prompt questionnaires described in \cite{yan2023robust}. These concepts are aggregated into a concept set $C = \{c_1, c_2, \cdots, c_N \}$ for each disease category. 
To extract image embeddings, we utilize the ChexAgent model \cite{chen2024chexagent}, a multimodal vision language model (VLM) tuned explicitly for CXR interpretation. ChexAgent outputs image embeddings $V \in \mathbb{R}^{H \times D}$, where $H$ is the image height, and $D$ is the embedding dimension. Large language models have demonstrated effectiveness in encoding clinical knowledge \cite{singhal2023large}. For each concept $c_i$, a text embedding $t_i \in \mathbb{R}^D$ is generated using the Mistral Embed Model \cite{jiang2023mistral}. This model is chosen for its efficiency and accuracy in embedding textual data into a high-dimensional vector space, suitable for subsequent similarity computations.

Given an image embedding $V$ and a set of concept embeddings $t_i$, we computed a similarity matrix $M_{i,j} \in \mathbb{R}^{H \times N}$ using cosine similarity,
where $i$ indexes over the image embedding pixels, and $j$ indexes over the concept embeddings. To reduce the dimensionality and focus on the most salient features, max pooling is applied to each similarity matrix $M_{i,j}$. This results in a singular value $s_i = \text{max}(M_i)$ for each concept, forming a concept vector $e= (s_1,s_2,\cdots,s_N)$. After that, the concept vector $e$ is normalized to a scale between 0 and 1 to maintain interpretability. A fully connected layer  $W_F \in \mathbb{R}^{M_c \times N}$ is then applied to classify the images into $M_c$ categories. The classification logits are computed as $z_i = W_F \cdot e_i$, where $z_i$ is the logit vector for the $i$-th image. Log-softmax is applied to the logits to obtain the log probabilities. The model is trained using categorical cross-entropy loss \cite{bishop2006pattern}. During inference steps, the model then provides the predicted class and associated concepts and their contributions, which are explicitly used in the report generation step.

\subsection{Robust Explanation-based Radiology Report Generation}
\label{subsec:rg}

We collected clinical documentation for each disease category from the National Institutes of Health (NIH). To facilitate efficient retrieval and analysis, embeddings for each document \( d_i \) were generated using the OpenAI embedding model \cite{brown2020language}, yielding vectors \( \mathbf{v}_i \in \mathbb{R}^n \) such that \( \mathbf{v}_i = f_{\text{embed}}(d_i) \), where \( f_{\text{embed}} \) is the embedding function. These embeddings were stored in Qdrant vector database \footnote{https://qdrant.tech/} \( \mathcal{Q} = \{ \mathbf{v}_i \mid i = 1, 2, \ldots, N \} \), with \( N \) as the number of documents. The multi-agent framework for Retrieval-Augmented Generation (RAG) includes specialized agents for each disease category, each implemented as a Reasoning and Acting (ReAct) agent \cite{yao2022react}.  For a classified disease category \( M_c \), the corresponding ReAct agent \( A_C \) is activated to retrieve relevant document embeddings \( D_C \) from \( \mathcal{Q} \), based on similarity to an input query \( q \): \( D_C = \{ \mathbf{e}_j \mid \text{sim}(\mathbf{v}_j, q) \geq \tau \} \), where \( \text{sim}(\cdot) \) is a similarity function and \( \tau \) a relevance threshold.

Alongside the ReAct agents, the framework incorporates a Radiologist Agent \( A_R \) and a Medical Writer Agent \( A_W \). The Radiologist Agent \( A_R \) uses the activated ReAct agent \( A_C \) as a tool to retrieve relevant clinical information from \( D_C \) and, based on identified concepts \( c_k \) from earlier stages, calculates an influence score \( s_k = \text{influence}(c_k, D_C) \) for each concept, where \( \text{influence}(\cdot) \) assesses relevance and diagnostic significance. These scores are assembled into a summary vector \( \mathbf{s} = (s_1, s_2, \ldots, s_m) \), with \( m \) as the number of relevant concepts. The Medical Writer Agent \( A_W \) receives this vector \( \mathbf{s} \) and composes a report by applying a generation function \( f_{\text{gen}} \), resulting in \( \text{report} = f_{\text{gen}}(\mathbf{s}) \). The final output \( y \) for a query \( q \) in category \( C \) can be represented as \( y = A_W(A_R(A_C(q, \mathcal{Q}))) \), where each agent sequentially processes and enhances the information to generate a detailed radiology report. This framework was implemented using CrewAI \footnote{https://www.crewai.com/} and facilitated by LlamaIndex \footnote{https://www.llamaindex.ai/}, ensuring efficient retrieval and high-quality report generation.

\section{Results and Discussion}
We evaluate the performance of both the interpretable classification using CBMs and the report generation module on the COVID-QU Dataset \cite{chowdhury2020can}, compiled by Qatar University, comprising 33,920 CXR images across three classes: COVID-19 (11,956 images), Pneumonia (11,263 images), and Normal (10,701 images).

% New Code for Side-by-Side Tables
\begin{table}[b]
    \centering
    \label{tab:side_by_side}
    \begin{minipage}{0.45\textwidth}
        \centering
        \caption{Classification Performance Comparision.} 
        \resizebox{\textwidth}{!}{
            \begin{tabular}{lcc}
\hline
\textbf{Model}                                   & \textbf{Covid-QU} & \textbf{Intepretability} \\ \hline
CLIP \cite{radford2021learning}                                            & 0.47              & No                      \\
Bio-VIL \cite{yan2023robust}                                        & 0.78              & No                       \\
Label-free CBM \cite{oikarinen2023label}                                  & 0.72              & Yes                      \\
Robust CBM \cite{yan2023robust} & 0.78              & Yes                      \\ \hline
Ours                                             & \textbf{0.81}     & Yes                      \\ \hline
\end{tabular}
        }
        \label{tab:comp}
    \end{minipage}%
    \hfill
    \begin{minipage}{0.45\textwidth}
        \centering
        \caption{Clustering Evaluation for Report Generation Approaches}
        \resizebox{\textwidth}{!}{
            \begin{tabular}{l@{\hspace{20pt}}c@{\hspace{20pt}}c@{\hspace{20pt}}c}
                \toprule
                \textbf{Metric} & \textbf{GPT4} & \textbf{Single Agent} & \textbf{Multi-Agent} \\
                \midrule
                Silhouette        & 0.37   & 0.41   & 0.27   \\
                Davies-Bouldin    & 1.11   & 0.96   & 1.44   \\
                Calinski-Harabasz & 69.94  & 93.99  & 44.78  \\
                Dunn              & 0.54   & 0.73   & 0.36   \\
                \bottomrule
            \end{tabular}
        }
        \label{tab:cluster}
    \end{minipage}
\end{table}

\begin{figure}[t]
    \centering
    \begin{subfigure}[b]{0.49\textwidth}
        \centering
        \includegraphics[width=0.65\textwidth]{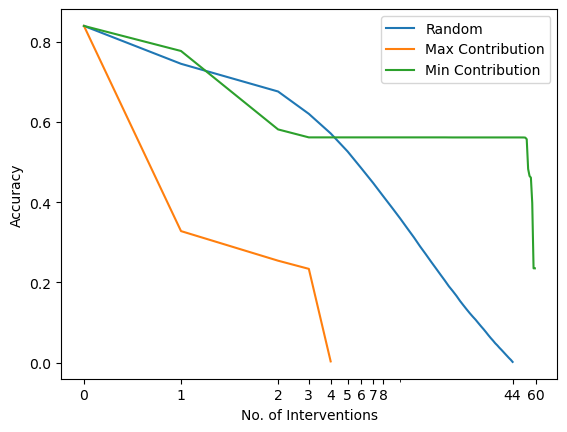}
        \caption{Model performance while removing concept features using different strategies.}
        \label{int_pred}
    \end{subfigure}
    \hfill
    \begin{subfigure}[b]{0.49\textwidth}
        \centering
        \includegraphics[width=0.65\textwidth]{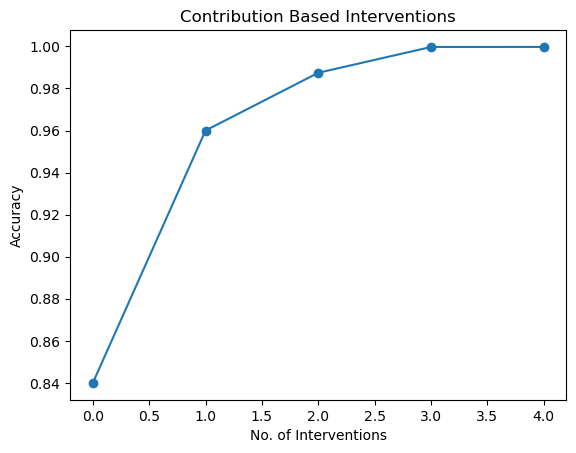}
        \caption{Model performance on intervening and correcting concepts for misclassified cases.}
        \label{int_corr}
    \end{subfigure}
    \caption{Evaluation of the Robustness of Concept set of the classification model.}
    \label{int}
\end{figure}

We compare the classification results with two types of baselines: visual encoders: CLIP \cite{radford2021learning}, Bio-VIL \cite{bannur2023learning}, and two other methods that are trained on CBMs: Label-free CBM\cite{oikarinen2023label} and Robust CBM \cite{yan2023robust}.
As shown in Table \ref{tab:comp}, Our classification model attains an accuracy of 81\% on the Covid-QU dataset. To further evaluate the interpretability and robustness of the concept contribution, we explored concept intervention techniques to correct the predictive output of the model \cite{shin2023closer,steinmann2023learning}. In Fig. \ref{int_corr}, we observe that correcting 3-4 concepts for misclassified test samples based on decreasing order in contribution scores yields a significant increase in performance. In Fig. \ref{int_pred}, the model performance on the test set is evaluated while removing the concept contributions in descending (Max Contribution), ascending (Min Contribution), and Randomly. The sharp decline in performance in the case of Max Contribution validates that the models indeed learn to predict from the concept contributions. The x-axis for Fig. \ref{int_pred} is in log-scale for better interpretation.

\begin{figure}[b]
    \centering
    \begin{subfigure}[b]{0.30\textwidth}
        \centering
        \includegraphics[width=\textwidth]{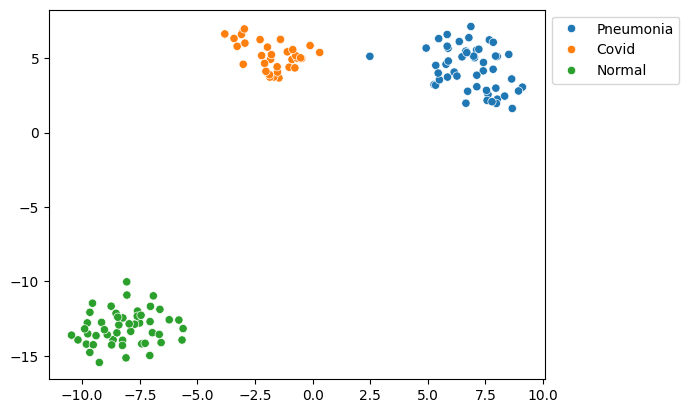}
        \caption{Multi Agent RAG}
        \label{clustersmulti}
    \end{subfigure}
    \hfill
    \begin{subfigure}[b]{0.30\textwidth}
        \centering
        \includegraphics[width=\textwidth]{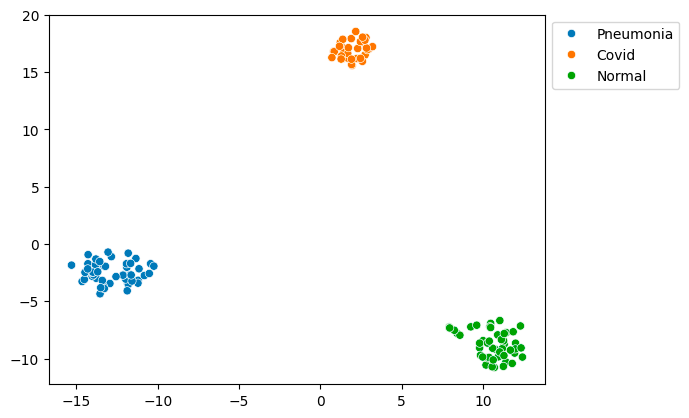}
        \caption{Single Agent RAG}
        \label{clusterssingle}
    \end{subfigure}
    \hfill
    \begin{subfigure}[b]{0.30\textwidth}
        \centering
        \includegraphics[width=\textwidth]{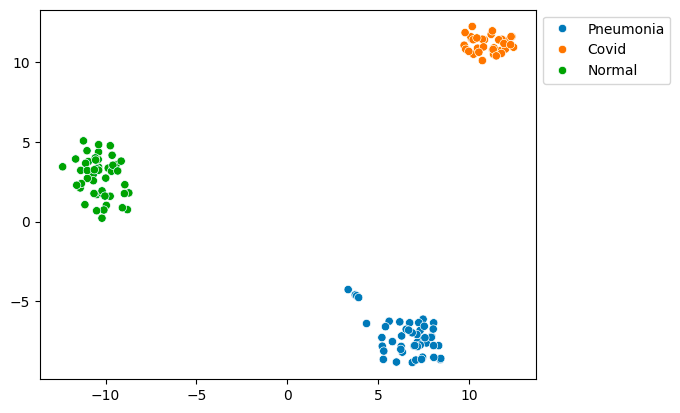}
        \caption{GPT 4}
        \label{clustersgpt}
    \end{subfigure}
    \caption{t-SNE visualization of the embeddings of generated reports}
    \label{clusters}
\end{figure}

To evaluate the effectiveness of our report generation process using the multi-agentic approach, we generate medical reports for all three classes: Pneumonia, COVID-19, and Normal. We compare the report generated with the same generated using GPT-4 and single-agent RAG, where one single agent is responsible for retrieving and generating reports from relevant documents. Each report is transformed into high-dimensional embeddings using the Mistral Embed Model, capturing the latent features. We apply t-distributed Stochastic Neighbor Embedding (t-SNE) \cite{van2008visualizing} to reduce the dimensionality for visualization and analysis. To quantify how well the reports are clustered for different diseases, we compute Silhouette Score \cite{rousseeuw1987silhouettes}, Davies-Bouldin Index, \cite{davies1979cluster}, Calinski-Harabasz Index \cite{calinski1974dendrite}, and Dunn Index \cite{dunn1974well}. The clustering evaluation metrics in Table \ref{tab:cluster} offer insights into the quality of clustering for the different approaches. While the Single Agent method achieves the highest Silhouette Score (0.41), lowest Davies-Bouldin Index (0.96), and highest Calinski-Harabasz Index (93.99), indicating the tightest and most distinct clusters, it sacrifices clinical interpretability. In contrast, the Multi-Agent approach has slightly lower metrics, such as a Silhouette Score of 0.27 and a Davies-Bouldin Index of 1.44, yet this method reflects the clinical reality more accurately. Specifically, the COVID-19 and Pneumonia clusters show some proximity in the Multi-Agent approach, as shown in the t-SNE plot in Fig. \ref{clusters}, which is medically justified due to the biological overlap between these conditions. Additionally, the Normal cluster remains well-separated, confirming the model's ability to differentiate fundamentally different health states. Thus, despite lower numerical scores, the Multi-Agent approach better captures the nuances required for effective medical report generation.

To further evaluate the generated reports, we employed LLM as a Judge \cite{zheng2023judging} approach, where five different LLMs — Llama 3.1 \cite{dubey2024llama}, Mistral \cite{jiang2023mistral}, Gemma2 \cite{team2024gemma}, LLaVA \cite{liu2024improved}, and GPT-3.5 Turbo \cite{brown2020language}, were used to assess the reports for Semantic Similarity, Accuracy, Correctness, Clinical Usefulness, and Consistency against a ground-truth reference generated by Dragonfly-Med \cite{chen2024dragonfly}, a multimodal biomedical visual-language model by Together AI, fine-tuned from Llama 3 that achieved state-of-the-art performance on several benchmarks. Table \ref{tab:combined} shows that using a multi-agent approach in report generation significantly enhances performance across all metrics compared to single-agent methods. For example, models like Mistral 7B exhibit notable improvements when employing the multi-agent approach, with Correctness increasing from 0.85 (Single Agent) to 0.95 (Multi-Agent) and Clinical Usefulness from 0.92 to 0.96. For further qualitative analysis, we use the Mixture of Agents (MoA) \cite{wang2024mixture} approach. For our evaluation, Llama3.1 \cite{dubey2024llama} and Mistral \cite{jiang2023mistral} act as proposer agents, while Medllama2 serves as the aggregator agent. The qualitative feedback from this process is used to perform a binary classification using GPT4-o, determining whether the generated report is clinically valid. As shown in the table below, the MoA method achieves scores of 0.81 with GPT-4, 0.82 with the Single Agent approach, and 0.85 with the Multi-Agent approach. The incremental improvement in scores—from 0.81 to 0.85—indicates that leveraging multiple agents enhances the quality and clinical validity of the generated reports, which aligns with the findings of \cite{li2024more}.

\begin{table*}[t]
  \caption{Evaluation of Report Generation Approaches using LLM as Judge}
  \label{tab:combined}
  \resizebox{\textwidth}{!}{
  \begin{tabular}{c|c@{\hspace{15pt}}c@{\hspace{15pt}}c|c@{\hspace{15pt}}c@{\hspace{15pt}}c|c@{\hspace{15pt}}c@{\hspace{15pt}}c|c@{\hspace{15pt}}c@{\hspace{15pt}}c|c@{\hspace{15pt}}c@{\hspace{15pt}}c}
    \toprule
    \textbf{Model} & \multicolumn{3}{c|}{\textbf{Semantic Similarity}} & \multicolumn{3}{c|}{\textbf{Accuracy}} & \multicolumn{3}{c|}{\textbf{Correctness}} & \multicolumn{3}{c|}{\textbf{Clinical Usefulness}} & \multicolumn{3}{c}{\textbf{Consistency}} \\
    \cmidrule{2-16}
    & \textbf{GPT4} & \textbf{Single} & \textbf{Multi} & \textbf{GPT4} & \textbf{Single} & \textbf{Multi} & \textbf{GPT4} & \textbf{Single} & \textbf{Multi} & \textbf{GPT4} & \textbf{Single} & \textbf{Multi} & \textbf{GPT4} & \textbf{Single} & \textbf{Multi} \\
    &  & \textbf{Agent} & \textbf{Agent} &  & \textbf{Agent} & \textbf{Agent} &  & \textbf{Agent} & \textbf{Agent} &  & \textbf{Agent} & \textbf{Agent} &  & \textbf{Agent} & \textbf{Agent} \\
    \midrule
    \textbf{Llama 3.1 8B} & 0.84 & 0.82 & 0.84 & 0.91 & 0.87 & 0.91 & 0.92 & 0.88 & 0.91 & 0.89 & 0.85 & 0.84 & 0.93 & 0.85 & 0.89 \\
    \textbf{Mistral 7B} & 0.79 & 0.88 & 0.89 & 0.84 & 0.88 & 0.94 & 0.85 & 0.85 & 0.95 & 0.88 & 0.92 & 0.96 & 0.86 & 0.88 & 0.96 \\
    \textbf{Gemma 2 9B} & 0.77 & 0.79 & 0.85 & 0.80 & 0.80 & 0.82 & 0.81 & 0.83 & 0.87 & 0.69 & 0.67 & 0.78 & 0.76 & 0.77 & 0.83 \\
    \textbf{LLaVA 9B} & 0.78 & 0.80 & 0.80 & 0.83 & 0.87 & 0.89 & 0.86 & 0.86 & 0.91 & 0.78 & 0.82 & 0.86 & 0.80 & 0.83 & 0.89 \\
    \textbf{GPT 3.5 Turbo} & 0.79 & 0.75 & 0.82 & 0.84 & 0.78 & 0.86 & 0.86 & 0.79 & 0.88 & 0.81 & 0.75 & 0.86 & 0.84 & 0.76 & 0.88 \\
    \midrule
    \textbf{Average} & 0.79 & 0.80 & 0.84 & 0.84 & 0.84 & 0.88 & 0.86 & 0.84 & 0.90 & 0.81 & 0.80 & 0.86 & 0.84 & 0.81 & 0.89 \\
    \bottomrule
  \end{tabular}
  }
\end{table*}

\section{Conclusion and Future Work}
This paper introduces an interpretable framework that integrates CBMs with a multi-agent RAG system for CXR classification and report generation. By capturing relationships between visual features and clinical concepts, our model provides competitive performance and clinically relevant explanations. The multi-agent RAG system enhances report quality and relevance, validated through evaluations by LLMs. This work bridges high-performing AI with the interpretability essential for clinical use, promising reliable AI-driven chest X-ray analysis. Future efforts will be to validate this approach to other imaging modalities and further refine the multi-agent system for adaptability and robustness.

\section*{Acknowledgments}
This work is funded by the Federal Ministry of Education, Science, Research and Technology (BMBF), Germany, under grant number 01IW23002 (No-IDLE).

\bibliographystyle{splncs04}
\bibliography{ref}

\end{document}